\documentclass[10pt,a4paper,twoside,twocolumn,american,prd,nofootinbib,superscriptaddress]{revtex4-1}
 
 \pdfoutput=1    
\usepackage{lmodern}
\usepackage[utf8]{inputenc}

\usepackage[T1]{fontenc}
\usepackage[utf8]{inputenc}
\usepackage{graphicx}
\usepackage{color}
\usepackage{babel}
\usepackage{natbib}
\usepackage{booktabs}
\usepackage{units}
\usepackage{amsmath}
\usepackage{amssymb}
\usepackage{esint}
\usepackage[normalem]{ulem}
\usepackage[unicode=true,pdfusetitle,
 bookmarks=true,bookmarksnumbered=false,bookmarksopen=false,
 breaklinks=false,pdfborder={0 0 0},backref=false,colorlinks=true]
 {hyperref}
\hypersetup{
 citecolor=blue,filecolor=blue,linkcolor=blue,urlcolor=blue}
\usepackage{subfigure} 
\usepackage{float}
\usepackage{verbatim}
\usepackage{xcolor}

\usepackage[utf8]{inputenc}



\def\be{\begin{equation}}
\def\ee{\end{equation}}
\def\bea{\begin{eqnarray}}
\def\eea{\end{eqnarray}}
\def\bes{\begin{split}}
\def\ees{\end{split}}

\def\kpc{{\rm kpc}}
\def\mpc{{\rm Mpc}}

\def\kpc{{\rm kpc}}
\def\cm{{\rm cm}}

\def\teq{t_{\rm eq}}

\def\rta{r_{\rm ta}}
\def\tta{t_{\rm ta}}
\def\rhom{\rho_{\rm max}}
\def\rcut{r_{\rm cut}}
\def\mpbh{M_{\rm PBH}}
\def\fpbh{f_{\rm PBH}}

\def\etal{{\it et al.~}}
\def\fw{f_{\rm WIMP}}

\begin{document}

\title{WIMPs and stellar-mass primordial black holes are incompatible}

\author{Julian Adamek}
\email{Julian.Adamek@qmul.ac.uk}
\affiliation{School of Physics \& Astronomy, Queen Mary University of London, 327 Mile End Road, London E1 4NS, United Kingdom}

\author{Christian T.~Byrnes}
\email{C.Byrnes@sussex.ac.uk}
\affiliation{Astronomy Centre, School of Mathematical and Physical Sciences, University of Sussex, Brighton BN1 9QH, United Kingdom}

\author{Mateja Gosenca}
\email{Mateja.Gosenca@auckland.ac.nz}
\affiliation{Department of Physics, University of Auckland, Private Bag 92019, Auckland, New Zealand}
\affiliation{Astronomy Centre, School of Mathematical and Physical Sciences,
University of Sussex, Brighton BN1 9QH, United Kingdom}

\author{Shaun Hotchkiss}
\email{S.Hotchkiss@auckland.ac.nz}
\affiliation{Department of Physics, University of Auckland, Private Bag 92019, Auckland, New Zealand}

\begin{abstract}
We recently showed that postulated
ultracompact minihalos with a steep density profile do not form 
in realistic simulations with enhanced initial perturbations. 
In this paper we assume that a small fraction of the dark matter consists of primordial black holes (PBHs) and simulate the formation of structures around them. We find that in this scenario halos with steep density profiles do form, consistent with theoretical predictions. 
If the rest of the dark matter consists of weakly interacting massive particles (WIMPs), we also show that WIMPs in the dense innermost part of halos surrounding the PBH would annihilate and produce a detectable
gamma-ray signal. The non-detection of this signal implies that PBHs make up at most one billionth of the dark matter, provided that their mass is greater than one millionth of the mass of the Sun.
Similarly, a detection of PBHs would imply that the remaining dark matter could not be WIMPs. 
\end{abstract}
\maketitle

\section{Introduction}

Interest in primordial black holes (PBHs) has increased dramatically since LIGO first detected black hole mergers \cite{Bird:2016dcv,Clesse:2016vqa,Sasaki:2016jop}. PBHs are an early universe relic which contain information about the initial density perturbations on very small scales. They are unique amongst all dark matter candidates since they are not a new particle. However, many different observational constraints exist, requiring that PBHs with masses comparable to a stellar mass can only make up a small fraction of the total dark matter. We primarily consider PBHs in the range of $1-100 \,\,M_\odot$ in this paper, but constraints which show that PBHs cannot constitute all of the dark matter exist for all PBHs with masses satisfying $\mpbh\gtrsim10^{-10}M_\odot$, see Fig.~5 of \cite{Niikura:2017zjd} for Subaru microlensing constraints and \cite{Sasaki:2018dmp} for a review of the many other observational constraints and their associated caveats.
Therefore, any constraint on the PBH abundance must also consider what the remaining dark matter is. 

Weakly interacting massive particles (WIMPs) are one of the most popular dark matter candidates. In this paper, we consider a mixed WIMP-PBH dark matter model, and show that the two cannot coexist (at least in the simplest case of WIMPs with the standard cross section, $\langle\sigma v\rangle=3\times10^{-26}\cm^3/\mathrm{s}$). If PBHs exist, they would have formed long before matter-radiation equality. Each PBH would then accrete an extremely dense halo of WIMPs around itself. The expected signal of gamma rays from WIMP annihilation in these high-density regions is incompatible with observational constraints, unless PBHs form less than about one billionth of the dark matter density.
To demonstrate this, we first need to calculate the dark matter density profile that forms around the PBHs.
Analytic work on high-density halos has been done previously, leading to the expectation that very steep halo profiles ($\rho(r)\propto r^{-9/4}$) develop in regions which were initially extremely dense. In our previous paper \cite{Gosenca:2017ybi} and in Delos \etal\cite{Delos:2017thv} it was shown that ultracompact minihaloes (UCMHs) with such steep density profiles do not form under realistic initial conditions.
Neither of these two papers included primordial black holes (PBHs).  

In this work we have performed the first simulations of UCMH formation in a universe containing a subdominant fraction of PBHs surrounded by particle dark matter (DM). Our PBH masses are of order ten solar masses. We show that the PBHs act as a strong gravitational seed, even during radiation domination, causing a halo of dark matter particles to form around each PBH. 
At small radii, the dark matter halos have a very large density, are spherically symmetric and have an $r^{-9/4}$ density profile. If the dark matter self-annihilates, e.g.~in the case of WIMPs, our profile would then be cut-off at very small radii due to this annihilation.
Our simulations are valid for any dark matter candidate, provided that it is cold on the scales probed by the simulation and particle-like in nature.

Our key conclusion is that a mixed dark matter model consisting of WIMPs and PBHs is excluded. If WIMPs were detected, then PBHs must be extremely rare, less than about one billionth of the dark matter density, unless they are very light with masses far below the reach of LIGO and VIRGO. Similarly, if any PBHs were detected then the remainder of the dark matter cannot be made out of WIMPs. 
This is because the WIMP annihilation from the extremely high-density regions close to PBHs would create a detectable signal of gamma rays. 

The incompatibility of  a mixed WIMP-PBH dark matter scenario was first discussed by Lacki and Beacom \cite{Lacki:2010zf}, then discussed in the context of PBH seeds for supermassive black holes by Kohri {\it et al.}~\cite{Kohri:2014lza} and recently discussed again by Eroshenko \cite{Eroshenko:2016yve} and Boucenna \etal \cite{Boucenna:2017ghj}. However, all of those papers relied on analytic estimates of the dark 
matter density profile. We know that the estimates for UCMH density profiles \emph{without a PBH seed} are far from realistic \cite{Gosenca:2017ybi,Delos:2017thv}. Therefore it is important to numerically check those with a PBH seed as well. Our numerical simulations put some of the analytic calculations onto a firm footing. We also contrast the somewhat contradictory results in \cite{Boucenna:2017ghj,Eroshenko:2016yve}. Lastly, we provide a very simple analytic estimate of the dark matter halo profile using a simplification of the methods of \cite{Boucenna:2017ghj,Eroshenko:2016yve}. We show that this matches the simulations well at small radii, where most of the WIMP annihilation occurs.

If one uses realistic UCMH density profiles, then the constraint on the amplitude of the primordial power spectrum arising from the non observation of WIMP annihilation is around $10^{-8}-10^{-6}$ \cite{Delos:2018ueo} (see also \cite{Nakama:2017qac}). An additional uncertainty in these constraints comes from the dependence on the assumed shape of the primordial power spectrum. This constraint is several orders of magnitude tighter than the $10^{-2}$ amplitude required to generate PBHs from large amplitude initial density perturbations.\footnote{This amplitude is subject to some uncertainties including the potential existence of primordial non-Gaussianity, the equation of state when the relevant scales reenter the horizon and uncertainties in the density threshold required to collapse and form a PBH. However, even taking into account these uncertainties the threshold remains orders of magnitude above the UCMH one. For more details see \cite{Byrnes:2018txb}.} However, the UCMH constraints only cover a specific range of scales. The smallest scale they probe depends on the WIMP mass and thermal velocity, but is typically between $10^5-10^7 \mpc^{-1}$ \cite{Bringmann:2011ut}. These scales correspond to horizon masses of $1-10^4 M_\odot$ and the resulting PBH is expected to form with comparable mass. UCMHs set no constraint on the  existence of PBHs with lower mass. PBHs may also form due to topological defects or other means that do not require an enhanced initial power spectrum. In these scenarios their formation is hence not affected by the UCMH constraints of \cite{Delos:2018ueo}, see the reviews \cite{Green:2014faa,Sasaki:2018dmp} and references therein.

One way to discriminate between astrophysical and primordial black holes would be a detection of black holes with mass below the Chandrasekhar mass. Their existence is motivated by the reduction in pressure during the QCD transition while the horizon mass is about one solar mass \cite{Byrnes:2018clq}, and they are detectable because LIGO is sensitive to compact objects with masses as low as $10^{-2}M_\odot$ \cite{Abbott:2018oah,Magee:2018opb}. 
The effective spin of the PBH pair before merging is another key observable, with LIGO and VIRGO finding most of the detected events are consistent with zero initial spin \cite{LIGOScientific:2018jsj}. This is expected for PBHs formed during radiation domination \cite{Chiba:2017rvs,Mirbabayi:2019uph}, but not those formed during an early matter dominated era \cite{Harada:2017fjm} and arguably harder to explain in astrophysical models \cite{Belczynski:2017gds,Piran:2018bbt}; see also \cite{Clesse:2017bsw}.

The plan of our paper is as follows: In Sec.~\ref{sec:profiles} we analytically derive the density profile around PBHs, showing that the thermal kinetic energy of the WIMPs has a negligible impact compared to the gravitational potential energy of stellar-mass PBHs. In Sec.~\ref{Sec:Hill} we study the stability of the dark matter haloes around PBHs at late times. In Sec.~\ref{sec:simulations} we simulate the halo formation for different choices of the initial power spectrum, before showing that PBHs and WIMPs are essentially incompatible in Sec.~\ref{sec:constraints}. We conclude in Sec.~\ref{sec:conclusions} and detail our analytical calculations in the appendices. 

\section{Halo profiles}\label{sec:profiles}

We consider a universe in which a small fraction of the dark matter is contained in PBHs and the rest is made up out of WIMPs. The black holes that LIGO has detected, or is sensitive to, have a mass in the range $10^{-2}M_\odot$ to $10^{2}M_\odot$ \cite{Abbott:2018oah,Magee:2018opb}. For this mass range a host of observational constraints suggest that PBHs must be subdominant to the rest of dark matter, i.e.~$\fpbh\equiv\Omega_{\rm PBH}/\Omega_{\rm DM}\ll1$ \cite{Sasaki:2018dmp}. There are caveats to this 
conclusion, for example uncertainties about the primordial and late time clustering of the PBHs and their mass function, with many constraints based on a monochromatic mass function, but there remains evidence that even when relaxing those unrealistic assumptions the constraints can still rule out $\fpbh=1$.

\subsection{The turnaround radius}

Using a Newtonian approximation, the physical distance $r$ between a test particle and a PBH of mass $\mpbh$ in an otherwise unperturbed Friedmann-Lema\^{\i}tre-Robertson-Walker universe changes as
\be \ddot{r}=-\frac{G\mpbh}{r^2}+\frac{\ddot{a}}{a} r \,,
\label{eq:r-eom}
\ee
where the first term is the Newtonian gravitational attraction and the second term is (during radiation domination) the \emph{deceleration} of the background expansion. In the absence of the PBH, $\dot{r}$ would always be positive. This is despite the fact that the contribution to $\ddot{r}$ from the background is still $<0$ (i.e. a deceleration). However, in a flat background the balance between inertia and deceleration is delicate, $\dot{r}$ asymptotically approaches zero but never reaches it. Therefore, once the Newtonian term from the PBH 
becomes larger in magnitude than the expansion term in equation \eqref{eq:r-eom} the particle decouples from the background expansion and very soon overcomes the outward inertia. This means that, to a reasonable approximation, the ``turnaround time'' (i.e. when $\dot{r}=0$ and the particles begins to move towards the PBH) can be obtained by equating the two terms in equation \eqref{eq:r-eom}. This is backed up by numerical results in appendix \ref{app:numprof}.

Using the acceleration equation, $\ddot{a}/a=-(1+3\omega)H^2/2$, we get a turnaround radius, $r=\rta$, defined by
\be G\mpbh=(1+3\omega) \frac{H^2}{2} \rta^3. \ee
We can gain some intuition from this. During radiation domination the total energy contained within a sphere of this radius is equal to half the mass of the PBH (setting the speed of light to $c=1$),
\be \frac12 \mpbh= \frac{4\pi}{3} \rho_{\rm tot} \rta^3. \ee
At matter-radiation equality the energy in the dark matter mass is equal to the energy in radiation. Therefore, at matter-radiation equality, the dark matter halo mass around a PBH is comparable to $\mpbh$, independent of the PBH mass. 

During radiation domination we can use $H=1/(2t)$ to calculate that
\be \rta \simeq \left(4G\mpbh t_{\rm ta}^2\right)^{1/3}. \qquad \text{(analytical estimate)}
\label{eq:ta-analytic}
\ee
$t_{\rm ta}$ is then the time that a shell is turning around at $\rta$. We present numerical solutions to  Eq.~\eqref{eq:r-eom} in appendix \ref{app:numprof}. These show that a much more accurate solution (to better than 0.1$\%$ accuracy) is reached by instead using
\be \label{eq:rta} 
\rta \simeq\left(2 G\mpbh t_{\rm ta}^2\right)^{1/3}, \qquad \text{(numerical estimate)}
\ee
so we will instead use this definition of $\rta$ for the duration of this paper.

\subsection{Kinetic and potential energy}
\label{sec:KE}

Our simulations initialise particles with zero thermal velocity. We are interested in their behaviour during radiation domination when dark matter is very sub-dominant. It might be expected that the thermal kinetic energies of the particles would have a measurable effect on the density profiles. We show in this section that for PBH masses of order $10 M_\odot$ this is not true.\footnote{In fact, if the dark matter 
mass satisfies $m_\chi \geq 100\, {\rm GeV}$, thermal kinetic energy can be ignored for any PBH mass $\gtrsim10^{-6}M_\odot$ - as we show in section \ref{sec:rcut}.} To do this we derive the ratio between the thermal kinetic energy and potential energy of a dark matter particle at turnaround and show that it is negligible. At any later time the ratio will be even smaller.

Extracting the gravitational potential at turnaround is straightforward. The dark matter particles have mass $m_\chi$, the PBH has mass $M_{\rm PBH}$ and their separation is the radius of turnaround, $r_{\rm ta}$. Thus
\begin{equation}
  E_p = \frac{GM_{\rm PBH} m_\chi}{r_{\rm ta}}.
\end{equation}

To know the kinetic energy at turnaround we need to scale the temperature of the dark matter when it decouples from the radiation down to its temperature at turnaround. The temperature of the dark matter drops proportionally to $1/a^2$. Note that this is different to the temperature of the Universe itself, which is dominated by radiation and thus drops proportionally to $1/a$.

The velocities of the dark matter are given by a Maxwell-Boltzmann distribution. To within a factor of a few, the peak, mean and rms of the velocity distribution are given by $mv^2 \simeq kT$. This means that if we use units where $k=1$ and $T$ is measured in eV, $E_k = T$. Therefore, in terms of the dark matter temperature at decoupling, $T_{\rm KD}$, and the time of decoupling, $t_{\rm KD}$, the kinetic energy at turnaround is
\begin{eqnarray}
\nonumber E_k & = & T_{\rm KD} \left(\frac{a_{\rm KD}}{a_{\rm ta}}\right)^2 \\
 \nonumber   & = & T_{\rm KD} \frac{t_{\rm KD}}{t_{\rm ta}} \\
    & = & \frac{T_{\rm KD} t_{\rm KD}\left(2GM_{\rm PBH}\right)^{1/2}}{r_{\rm ta}^{3/2}}\,.
\end{eqnarray}

The ratio between kinetic and potential energy can now be expressed as
\begin{equation}\label{eq:KEvsPE}
  \frac{E_k}{E_p} = \left(\frac{T_{\rm KD}}{m_\chi}\right) \left(\frac{t_{\rm KD}}{\sqrt{r_{\rm ta}}}\right) \left(\frac{2}{GM_{\rm PBH}}\right)^{\frac{1}{2}}.
\end{equation}

To explore this ratio as a function of $r_{\rm ta}$ and $M_{\rm PBH}$ we first need to choose a dark matter model to give us $T_{\rm KD}$, $t_{\rm KD}$ and $m_\chi$. For this we follow \cite{Boucenna:2017ghj}. Specifically, we take the temperature of kinetic decoupling to be given by equation (B5) in \cite{Boucenna:2017ghj},
\be
T_{\rm KD} = \frac{m_{\chi}}{\Gamma(3/4)} \left( \frac{\alpha\, m_\chi}{M_{\rm Pl}}\right)^{1/4}\,,
\ee
with
\be
\alpha \equiv \sqrt{\frac{16\pi^3 g_\star(T)}{45}}\,,
\ee
and $g_\star = 61.75$ from equation (4) of \cite{Boucenna:2017ghj} and the text below it. The time at decoupling is then found from the Friedmann equation
\be
\frac{1}{2t} = \frac{\alpha T^2}{M_{\rm Pl}},
\ee
which is equation (3) in \cite{Boucenna:2017ghj}. Finally, the dark matter particle mass is taken to be $m_\chi = 100$ GeV, again following \cite{Boucenna:2017ghj}. We note that if we were to follow the procedure in \cite{Eroshenko:2016yve} instead our results would be very similar, with small changes due to small differences in the particle physics model underlying the dark matter. In both cases, for LIGO-like PBH masses, the kinetic energy is at least 100 times smaller than the potential energy at the turnaround time for all relevant radii.

\begin{figure}
	\begin{flushleft}
		\includegraphics[scale=0.6]{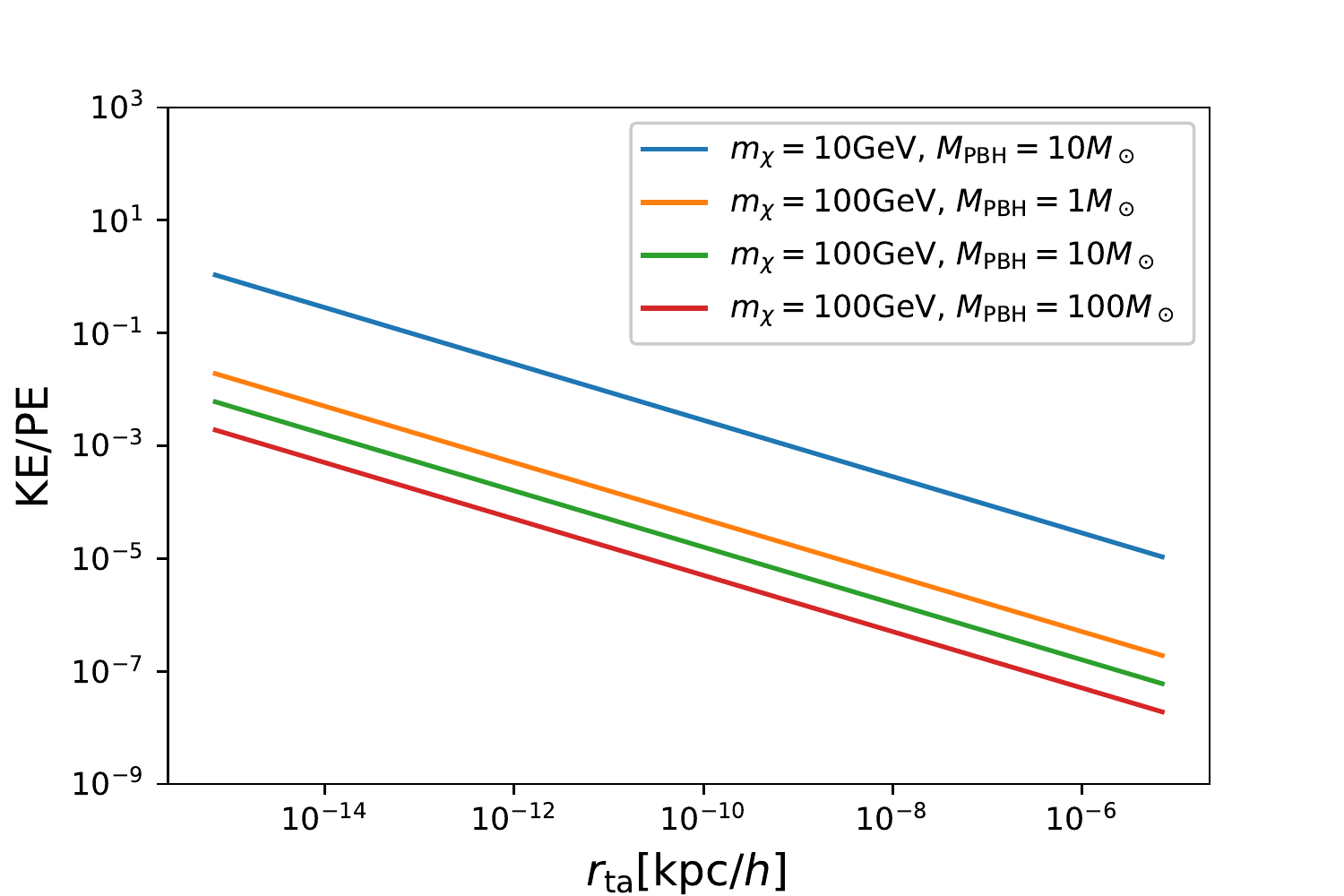}
	\end{flushleft}
	\caption{Ratio of the thermal kinetic energy to the potential energy at the turnaround radius for various PBH masses.}
	\label{fig:KEvsPE}
\end{figure}

In figure \ref{fig:KEvsPE} we show this ratio for three different PBH masses. We have also included one line for a lighter dark matter particle mass to show that kinetic energy can be relevant for sufficiently light dark matter at sufficiently small radii. However, for the $100$ GeV mass dark matter we see that even for a $1 M_\odot$ mass PBH the kinetic energy is negligible for all the radii plotted. We will see in section \ref{sec:rcut} that this plot is very conservative because we are actually only interested in radii above $r_{\rm ta} \sim 10^{-7} {\rm kpc}/h$ (for LIGO-like PBH masses). This is because dark matter annihilation would reduce the density observable today at radii smaller than this.

\subsection{The analytical density profile}

We can make a simple estimate of the density profiles surrounding a PBH by assuming the particles are frozen in at turn-around with their density matching the background density at that time.

During radiation domination, the resulting density profile is
\be 
\begin{split}
\rho_{\rm DM}(r)&=\rho(r(\tta)) \\
&=\frac{\rho_{\rm eq}}{2} \left(\frac{a}{a_{\rm eq}}\right)^{-3} \\
&\simeq \frac{\rho_{\rm eq}}{2} \left(\frac{t}{t_{\rm eq}}\right)^{-3/2} \\
&\simeq \left(\frac{\rho_{\rm eq}}{2}\right) t_{\rm eq}^{3/2} \left(2 G \mpbh\right)^{3/4} r^{-9/4}\,,
\label{eq:profile}
\end{split}
\ee
where $\rho_{\rm eq}$ is the density of the Universe at matter-radiation equality. This density is twice the background density of the matter at this time, hence the extra factor of one half. Note that this generates a steep $r^{-9/4}$ profile. This was the profile derived for the spherically-symmetric collapse in an Einstein-de Sitter universe \cite{1984ApJ...281....1F, 1985ApJS...58...39B} and expected for UCMHs \cite{Bringmann:2011ut} (see also \cite{Berezinsky:2013fxa,Berezinsky:2014wya}). Given that most particles will later spend some time closer to the PBH than their turnaround radius but that very few will move further away the actual profile will necessarily be even more compact. In fact, in appendix~\ref{app:halo} we show that including the dynamics of the nearly radial orbits of the particles after turn-around does not change the profile shape but increases the density by about $50\%$. Phenomena that could alter this conclusion would be the halo being disrupted later on, or thermal kinetic energy being relevant at turnaround. We have addressed when kinetic energy is relevant in section \ref{sec:KE} and we address the possibility of halo disruption in section \ref{Sec:Hill}.

\subsection{WIMP annihilation and the maximum density}
\label{sec:rcut}

There is a maximum possible WIMP density today due to their self-annihilation \cite{Bringmann:2011ut}
\begin{equation}  \label{eq:rho-max} 
\begin{split}
\rhom&=\frac{m_\chi}{\langle\sigma v\rangle t_0}  \\
&\simeq \left(\frac{m_\chi}{100{\rm GeV}}\right)\left(\frac{3\times10^{-26}\cm^3/\mathrm{s}}{\langle\sigma v\rangle}\right)\left(\frac{4\times 10^{17}\mathrm{s}}{t_0}\right)\\
&\; \times 1.5\times 10^{-14} \mathrm{g}/\cm^3, 
\end{split}
\end{equation}
where $t_0$ is the age of the Universe, $m_\chi$ is the WIMP mass and $\langle\sigma v\rangle$ is the WIMP cross section. The WIMP annihilation cross section is chosen such that WIMPs form most of the dark matter and we neglect any velocity dependence of this thermally averaged quantity. For the reference values shown in the equation above this corresponds to a density contrast today of $1+\delta_0=\rhom/\rho_0\simeq10^{16}$. We note that this maximum density was not derived for particles undergoing radial motion. See \cite{Lacki:2010zf} for a discussion of the WIMP survival time inside dark matter haloes. 

The constant value of the density extends up to some radius $r_{\rm cut}$, where the power-law profile begins. We estimate this radius by equating Eq.~\eqref{eq:profile} with \eqref{eq:rho-max} to be 
\be 
\begin{split}
\label{Eq:rcut}
\rcut &= \left(\frac{\rho_{\rm eq}}{2\rhom}\right)^{4/9} \left(2 G \mpbh \: \teq^2 \right)^{1/3} \\
&\simeq \left(\frac{m_\chi}{100{\rm GeV}} \right)^{-4/9}\left(\frac{\mpbh}{M_\odot}\right)^{1/3} 
1.3 
\times10^{-7}  \kpc \; h^{-1}.
\end{split}
\ee
On the second line we have assumed the WIMP parameters shown in Eq.~\eqref{eq:rho-max} and used $\teq=2.4\times10^{12}\rm{s}$, $\rho_{\rm eq}=2.1\times10^{-19} \rm{g}/\cm^3$, and $h=0.7$.

\begin{figure}
	\begin{flushleft}
		\includegraphics[scale=0.6]{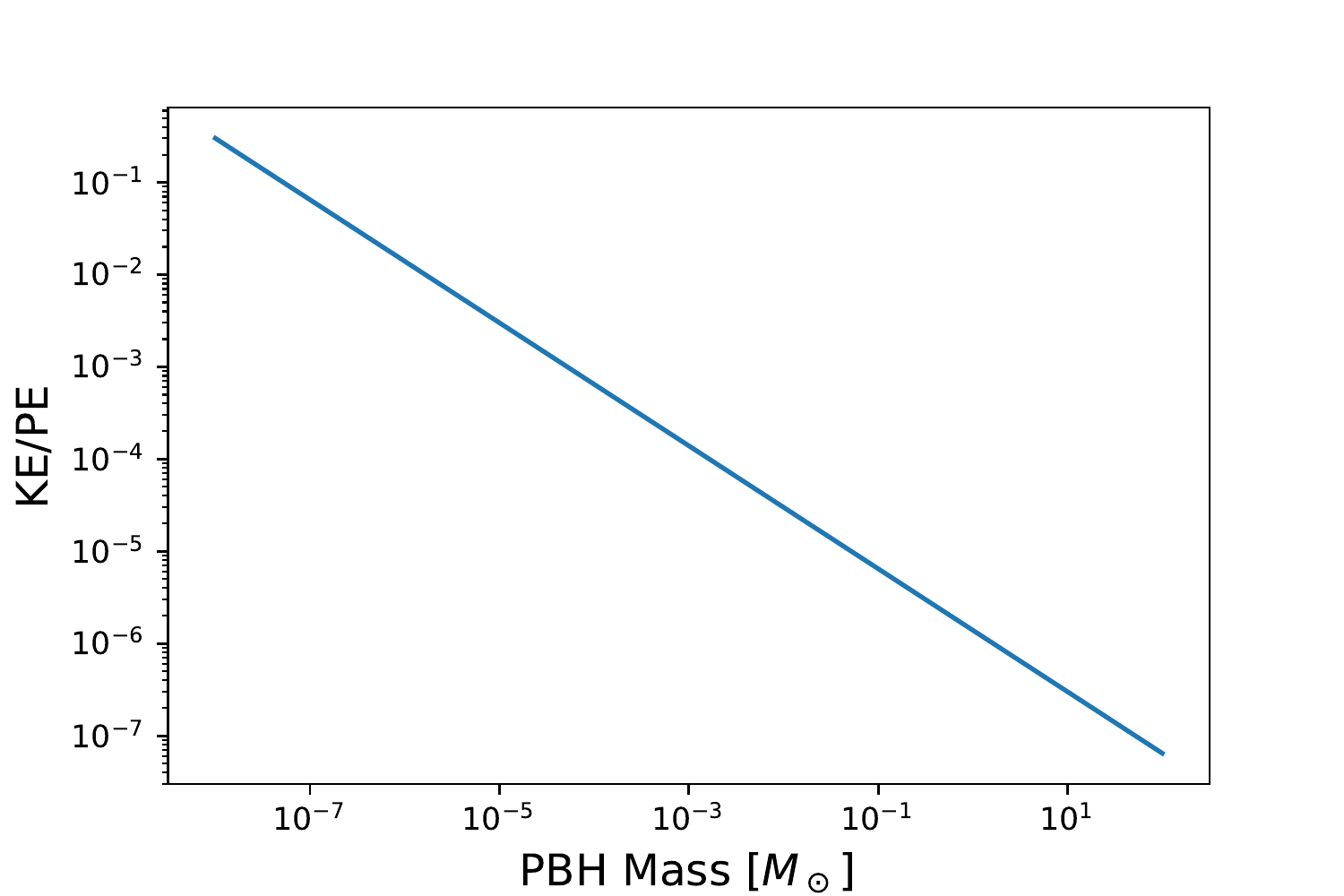}
	\end{flushleft}
	\caption{Ratio of the thermal kinetic energy to potential energy at $\rcut$ given by \eqref{Eq:rcut} plotted for various PBH masses. For this plot we have used a WIMP mass of $m_\chi = 100 \,{\rm GeV}$.}
	\label{fig:KEvsPE_rcut}
\end{figure}

In figure \ref{fig:KEvsPE_rcut} we use equation \eqref{eq:KEvsPE} to show the ratio of thermal kinetic energy to potential energy at $\rcut$ (calculated at the time when $\rcut$ is the turnaround radius). This is plotted as a function of the seed PBH mass. We see that thermal kinetic energy is not an important factor at radii $\gtrsim \rcut$ for PBH masses $\gtrsim 10^{-6} M_\odot$ (for all dark matter masses $m_\chi \geq 100 \,{\rm GeV}$).

\subsection{UCMH stability}
\label{Sec:Hill}

As we have demonstrated, the relevant WIMP halo forms around a PBH before matter-radiation equality, but the WIMP annihilation signal is only (potentially) detectable from nearby halos at $z=0$. Therefore it is crucial to know whether the halos are disrupted at any time between their formation and today \cite{Stref:2016uzb}.

For $\fpbh \ll 1$ and PBHs of the order of $10 M_\odot$ the most likely cause of disruption of the UCMH is a close encounter with a star. 
We are mostly interested in the disruption of the UCMH profile at distances close to $\rcut$ 
because most of the annihilation signal comes from that region (e.g.~$2/3$ of the signal comes from within $\sim 1.6~\rcut$). Such disruption only occurs if the distance of closest approach is of the order of $\rcut$. 
The rate of such encounters can be estimated as
\be
Z \simeq v \pi \rcut^2 n_\mathrm{star}\,,
\ee
where $n_\mathrm{star}$ is the number density of stars and $v$ is the typical relative velocity in encounters. For $\rcut \sim 10^{-7} \kpc \; h^{-1}$, and using $v \sim 250$ km s$^{-1}$ and $n_\mathrm{star} \sim 0.1 \; \mathrm{pc}^{-3}$ as typical for the Milky Way disk, we obtain $Z \sim 0.1$ Gyr$^{-1}$. Outside of the galactic disk the rate of disruptive encounters is negligible. We therefore estimate that UCMHs that spend most of the time outside the Milky Way disk in the halo are unlikely to be disrupted. Numerical simulations furthermore show that close encounters with stars do not always lead to a complete disruption \cite{Ishiyama:2010es}.

Due to the ultra-compactness of the UCMHs, tidal disruption in the mean-field potential of the Milky Way is not an issue. To show this explicitly, we can compute the tidal radius \cite{King:1962wi}
\be
r_\mathrm{tidal} \simeq R \left( \frac{\mpbh}{M_\mathrm{MW}} \right)^{1/3}\,,
\ee
where $R$ is the perigalactic distance of the UCMH and $M_\mathrm{MW}$ is the mass of the Milky Way. Clearly, $r_\mathrm{tidal} \gg \rcut$ unless the UCMH orbits extremely close to the galactic center -- in which case close encounters with stars become more likely as well.

\section{Numerical simulations}\label{sec:simulations}

We performed N-body simulations of the halo forming around a PBH during radiation and matter dominated eras. For this purpose we modified the publicly available code {\sc GADGET-2} to account for the radiation in the background expansion. We simulated an $(8 \kpc /h )^3$ volume with $256^3$ particles. 

Our particle template includes a PBH seed of $30~M_\odot$ that is displaced along with the dark matter particles, i.e.\ it starts at rest relative to the local matter. We do not adjust the initial particle velocities for accretion, but instead allow for a sufficiently long relaxation period so that the decaying modes damp out and the accretion can reach a steady state. We simulate for two types of initial perturbations:

\begin{itemize}
 \item A primordial power spectrum that is extrapolated from the cosmic microwave background (CMB) all the way down to the scales of our simulation assuming no running of the spectral index. The amplitude $A_s = 2.215 \times 10^{-9}$ at the CMB pivot scale of $0.05~\mathrm{Mpc}^{-1}$ and the spectral index $n_s = 0.9619$ are chosen to be compatible with current CMB constraints. The amplitude of perturbations on the scales of $\kpc$ is extremely small in this case, making it very hard to explain PBH formation in the first place. This simulation starts at an initial redshift of $z = 100\,000$. 

 \item A primordial power spectrum that is modified at small scales in order to allow for significant PBH formation in the $30~M_\odot$ mass range. On the scales resolved by our simulation we assume a power law with a blue tilt of $n_s-1 = 2$. This spectral index is half the steepest possible within the context of canonical single-field inflation \cite{Byrnes:2018txb}. The amplitude is chosen sufficiently high to make the PBH formation plausible. Extrapolated back to the CMB pivot scale it corresponds to $A_s = 2.75 \times 10^{-17}$. This number should not be compared to CMB constraints as any plausible scenario would require significant running of the spectral index. However, we do not need to specify the shape of the primordial power spectrum outside of our dynamical range. Since the perturbations at the $\kpc$ scale are much larger than in the previous case, we start this simulation at a higher initial redshift of $z = 5\,000\,000$ in order to justify using a linear description.

\end{itemize}

\begin{figure}
	\begin{center}
		\includegraphics[scale=0.5]{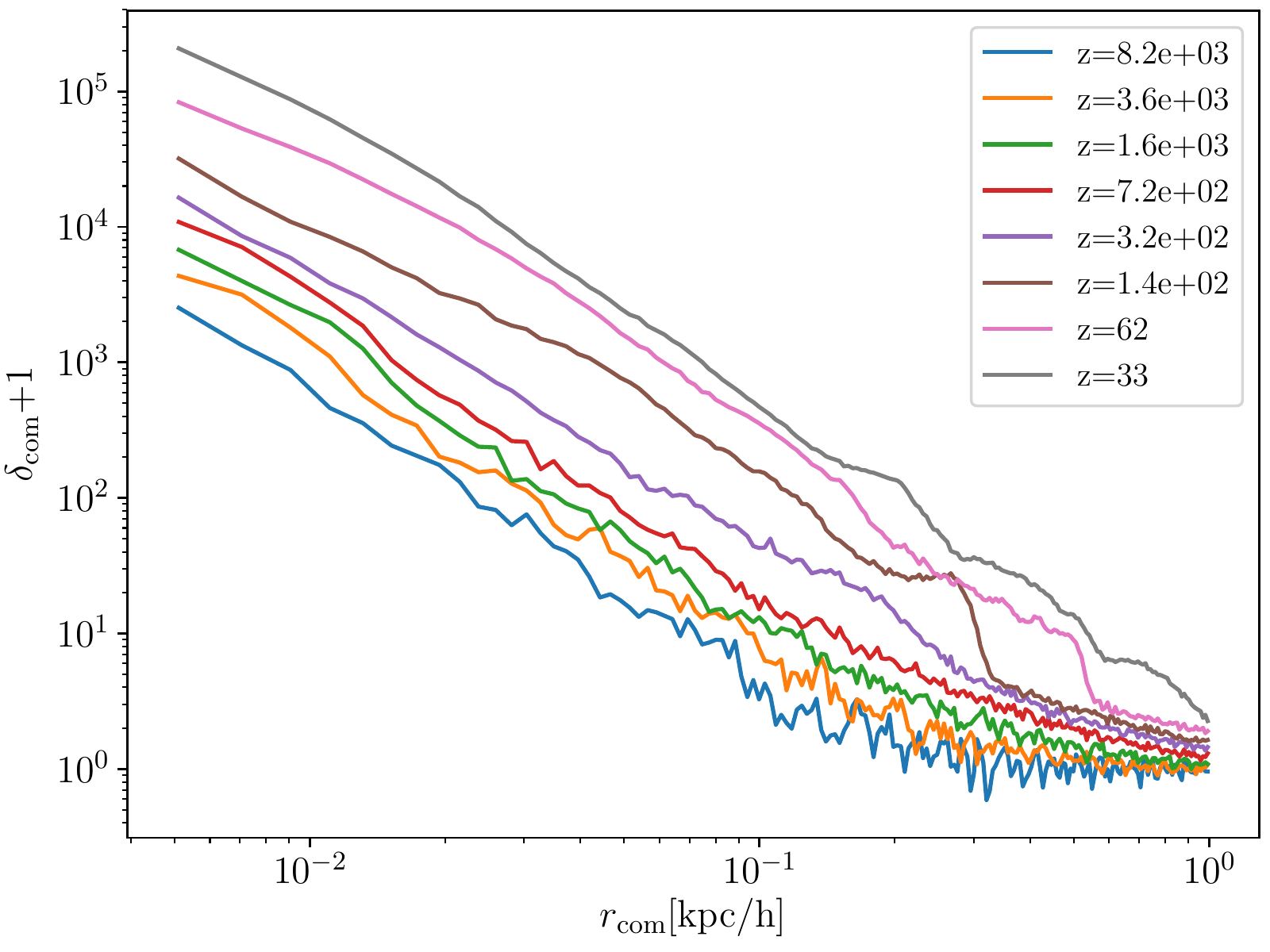}
	\end{center}
	\caption{The comoving density profile of a halo forming around a $30 M_{\odot}$ PBH in a $\Lambda$CDM background for different redshifts. Initially, the profile forms as a single power-law. As the surrounding material accretes onto the halo, its profile becomes more and more concave.} 
	\label{fig:single:comoving:256:12}
\end{figure}

\begin{figure}
	\begin{center}
		\includegraphics[scale=0.5]{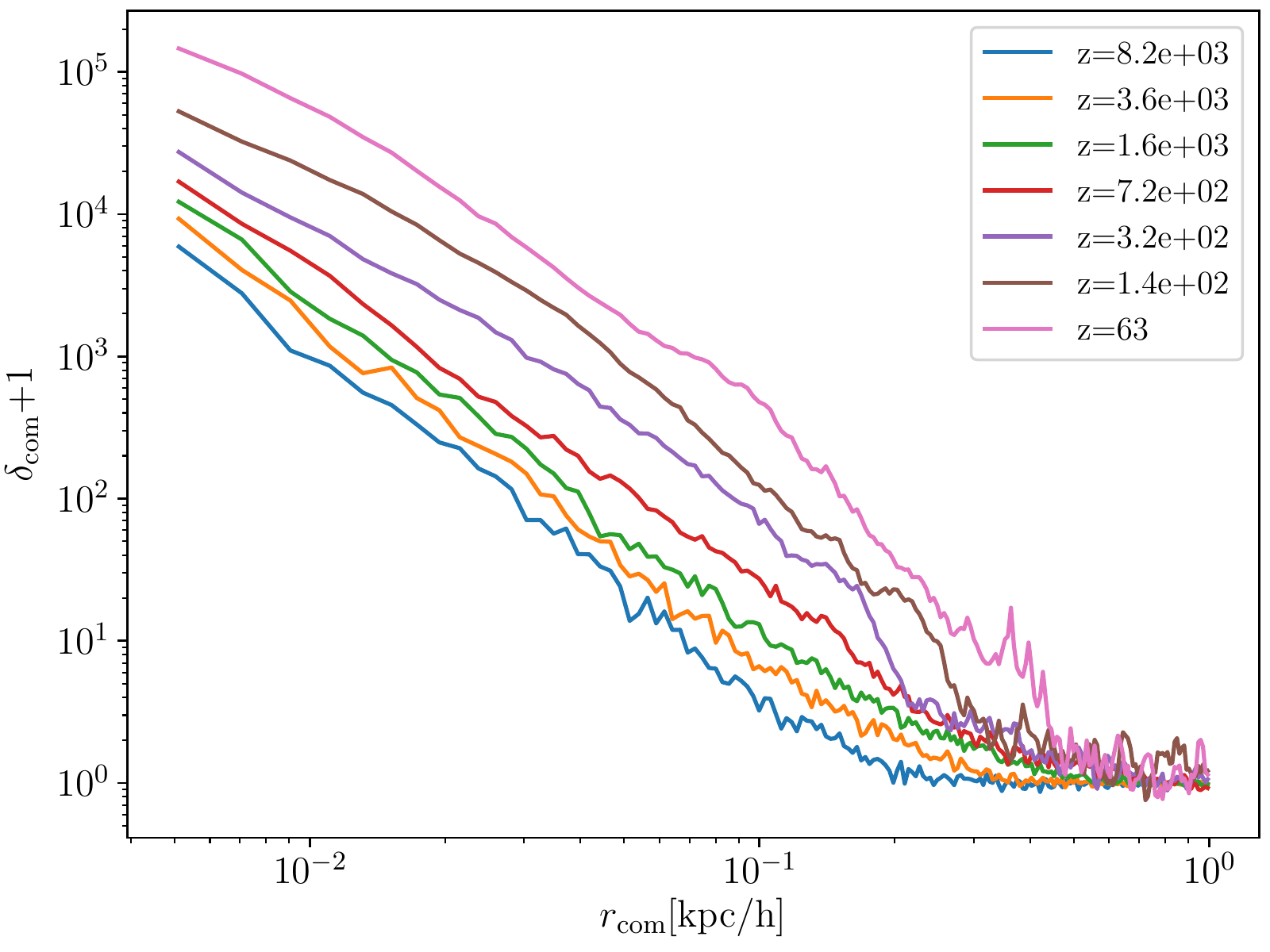}
	\end{center}
	\caption{The same as Fig. \ref{fig:single:comoving:256:12} but with the enhanced power spectrum, with spectral index $n_s-1=2$. } 
	\label{fig:boostedBG:comoving:256:12}
\end{figure}

In Figures \ref{fig:single:comoving:256:12} and \ref{fig:boostedBG:comoving:256:12} we show the profiles in comoving coordinates around the PBH for a $\Lambda$CDM universe and the one with a boosted power spectrum, respectively. The density spike caused by the PBH is excluded from the plots. 

\begin{figure}
	\begin{center}
		\includegraphics[scale=0.5]{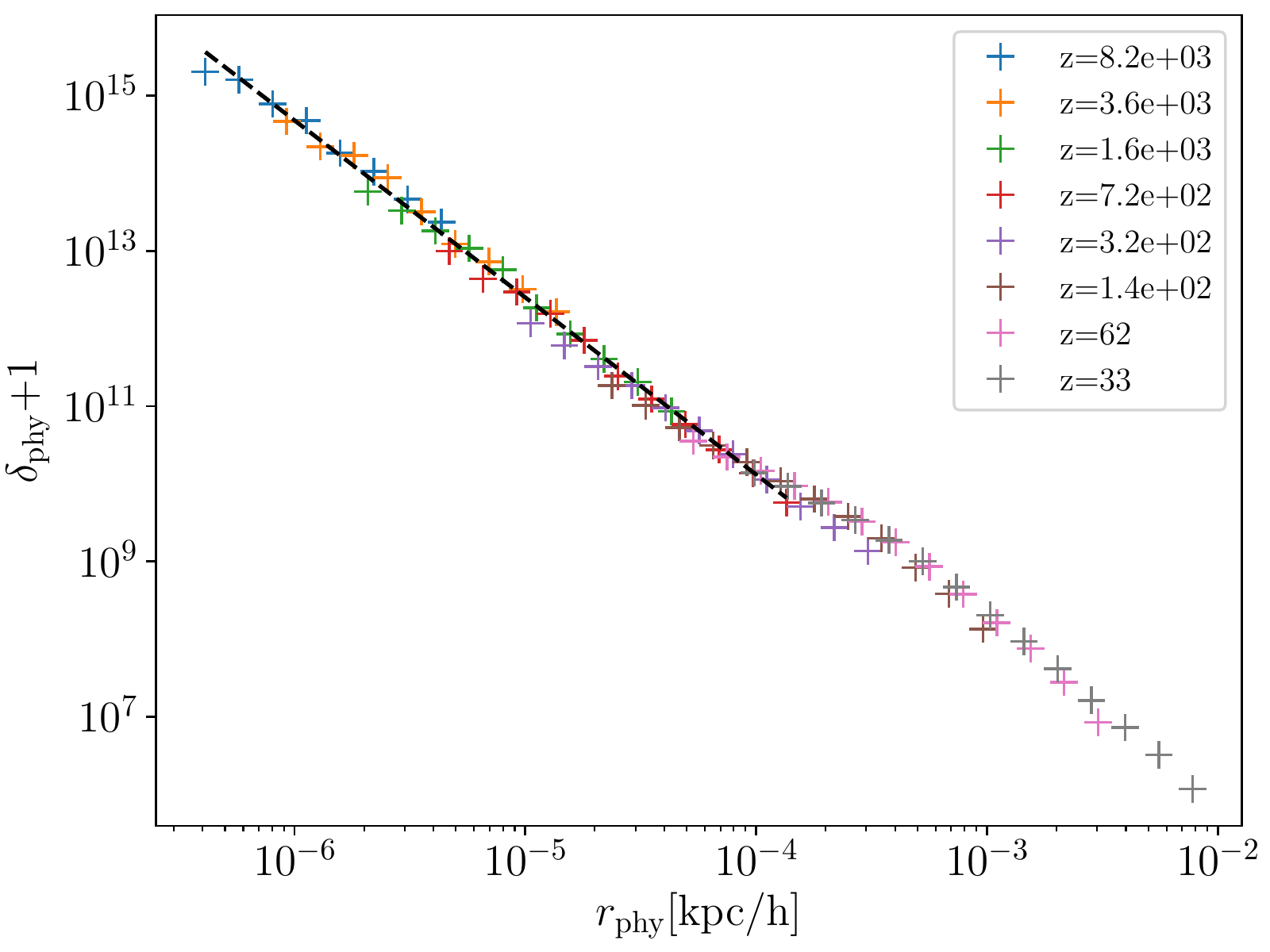}
	\end{center}
	\caption{The profile of a halo around a PBH in physical coordinates. The four inner most profiles were fit with a power-law profile: $\alpha =2.28$,  and $C=2.5 \times 10^{12}$.} 
	\label{fig:single:comoving:256:12:logr}
\end{figure}

\begin{figure}
	\begin{center}
		\includegraphics[scale=0.5]{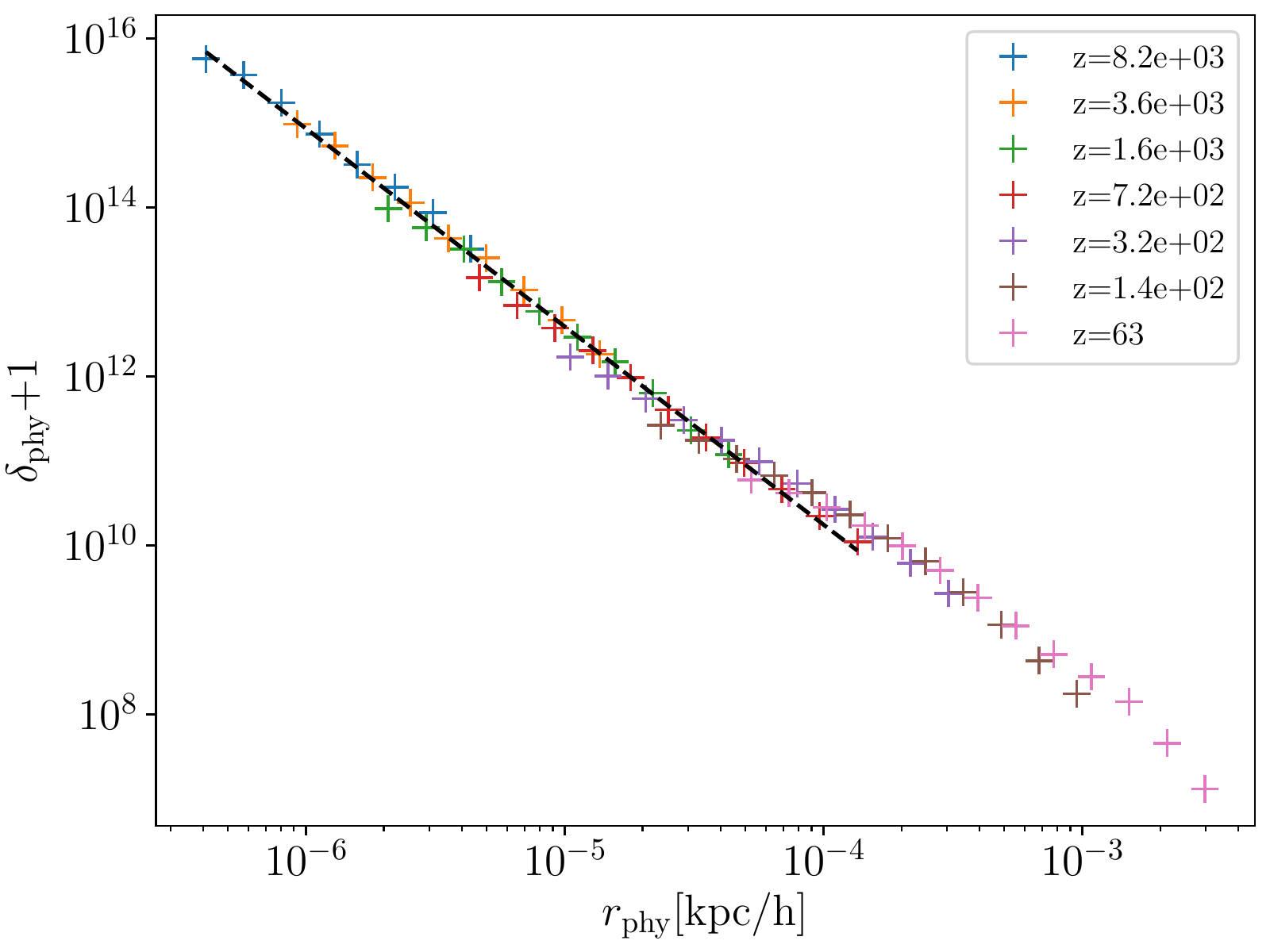}
	\end{center}
	\caption{The physical density profile of the halo for the boosted power spectrum, with spectral index $n_s-1=2$. The parameters of the fit are $\alpha =2.35$,  and $C=3.9 \times 10^{12}$.} 
	\label{fig:boostedBG:comoving:256:12:logr}
\end{figure}

In Figures \ref{fig:single:comoving:256:12:logr} and \ref{fig:boostedBG:comoving:256:12:logr} we show the profiles in physical coordinates. 
Here, each profile is shown between $r \sim2 \epsilon $ and $r \sim r_{\mathrm{ta}}/2$ where $\epsilon$ is the softening length in the simulation and $r_{\mathrm{ta}}$ is the turn-around radius. 
The softening length determines the scale below which the force between two simulation particles is suppressed so that it does not diverge when the separation between the particles approaches zero. 
In our simulations it was set to $\epsilon_{\mathrm{com}} \sim 1.95 \times 10^{-3} \kpc /h $ and therefore the smallest scale we resolve at the redshift of the earliest output is $r^{\min}_{\mathrm{phy}}= 4.8 \times 10^{-7} \kpc/h$, roughly the same as $\rcut$ given in Eq.~\eqref{Eq:rcut}. Following arguments laid out in \cite{Power:2002sw} we estimate that our choice of softening length ensures that strong discreteness effects do not occur within a comoving distance of $\sim 0.17 \kpc /h$ from the PBH, which is sufficient for our analysis.

The halo first forms as a single power-law function. As the simulation progresses, a Navarro-Frenk-White (NFW)-like profile is accreted at greater radii, but the steep profile in its interior remains intact (see the discussion in Section \ref{Sec:Hill} for the justification). 

We fit the interior of the density profile with a power law 
\be 
\delta_{\mathrm{phy}}+1 = C \: \left( \frac{r}{r_0} \right)^{-\alpha} \,,
\ee
where $\alpha$ and $C$ are the two parameters of the fit and the pivot scale $r_0 =10^{-5} \kpc ~ h^{-1}$ is chosen approximately in the middle of our fitting range. 
The fitting with the power-law was performed on the density profiles obtained from the four earliest snapshots which correspond to the central region of the profile before the NFW-like profile. 
In both cases we find $\alpha$ to be close to $9/4$. The profile of the halo embedded in the enhanced background is slightly steeper and reaches somewhat higher density in the centre, but it nevertheless forms in a very similar way.
This shows that even if the power spectrum were enhanced on scales that ensure the formation of $30~M_\odot$ PBHs, the surrounding enhanced perturbations would not have a significant effect on the formation of the dark matter halo around the PBH. 

\begin{figure}
	\begin{flushleft}
		\includegraphics[scale=0.49]{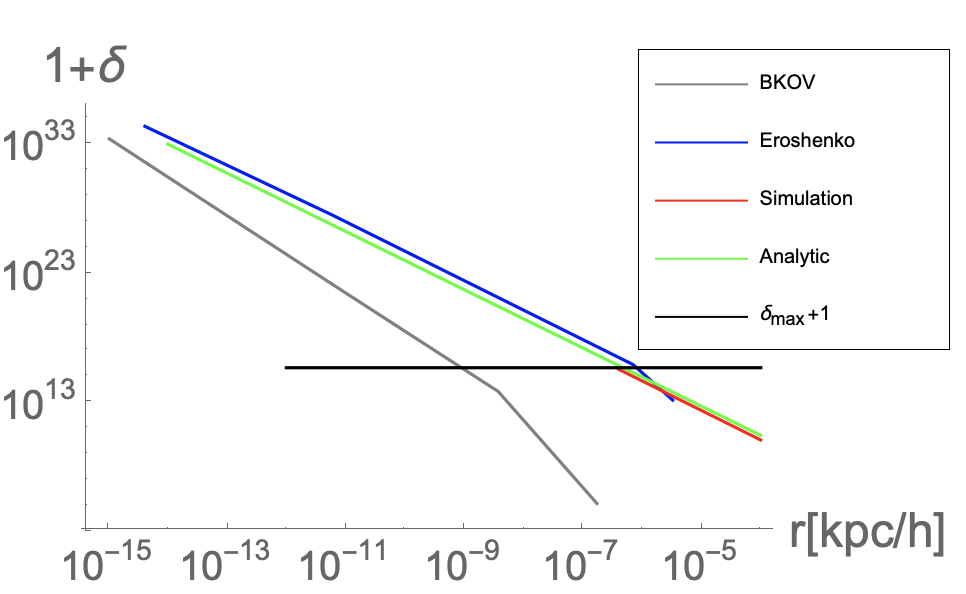}
	\end{flushleft}
	\caption{The density profiles of Boucenna {\it et al.}~\cite{Boucenna:2017ghj}, labelled BKOV, Eroshenko \cite{Eroshenko:2016yve}, our analytic estimate \eqref{eq:profile}, our simulation result taking the best-fit paramaters from Fig.~\ref{fig:single:comoving:256:12:logr} and the maximum density contrast today. Note that the simulation is for a 30$M_\odot$ black hole whilst the other three profiles are derived for a 10$M_\odot$ black hole.}
	\label{fig:dec-profile}
\end{figure}

Figure \ref{fig:dec-profile} shows the density profile we derived analytically, the profile generated by our simulation as well as those shown in \cite{Boucenna:2017ghj} and \cite{Eroshenko:2016yve}. 
All profiles are chosen for a $10M_\odot$ black hole apart from the simulated one which is three times more massive. Notice how the simulated and analytic density profiles are similar to each other and the result in \cite{Eroshenko:2016yve} but orders of magnitude more dense than the result in \cite{Boucenna:2017ghj}. The horizontal line shows the maximum possible density at late times, given by \eqref{eq:rho-max}. 
The simulation result is only plotted 
down to the radius
which we can numerically resolve, 
roughly
$r_{\rm cut}$, where the density profile reaches the maximum density at late times.   

\section{Constraints on the PBH fraction of dark matter}\label{sec:constraints}

Here we follow \cite{Boucenna:2017ghj} whose reasoning is as follows. In a scenario where the PBH seeds all have similar masses, and the UCMHs are likewise similar, they will all have similar absolute gamma-ray luminosities. To the extent that these macroscopic dark matter structures trace the overall distribution of dark matter well, i.e.\ without significant bias above a sufficiently large coarse-graining scale, their diffuse emission is directly proportional to the coarse-grained dark matter density. The same is true in a completely unrelated scenario where there are no PBHs and all of the dark matter is made up of particles that undergo one-body decay into gamma radiation. Coarse graining is an important step for making this analogy, since for resolved PBHs the gamma-ray emission depends quadratically on the local WIMP density due to the two-body nature of the annihilation process.

For the decaying dark matter scenario, the observed diffuse gamma-ray background has been used to put constraints on the unknown decay rate, where one typically assumes that the decaying species accounts for all of dark matter. Conversely, if one knew the decay rate, one would obtain a constraint on the abundance of the decaying species. Using the above analogy we can therefore obtain a bound on the PBH fraction $f_{\mathrm{PBH}}$ for any given PBH mass once we assume a complete model for the gamma-ray luminosity of these objects,
\be
f_{\mathrm{PBH}} = \frac{\Gamma_{\mathrm{DM}} \: M_{\mathrm{PBH}}}{\Gamma_{\mathrm{PBH}} \: m_\chi }.
\label{fPBH}
\ee
For $\Gamma_{\mathrm{DM}}$ we use (consistent with \cite{Boucenna:2017ghj}) constraints from \cite{Ando:2015qda}. Their Fig.~(3.f) shows that the life time of dark matter particles is greater than $\tau_{\mathrm{DM}} = \Gamma_{\mathrm{DM}} ^{-1} \gtrsim 10^{28} s$, at least in the range of the dark matter particles masses to which the experiment is sensitive: $ 10 \mathrm{GeV} < m_{\chi} < 10^{4} \mathrm{GeV}$. 

The WIMP annihilation signal is obtained as 
\be 
\Gamma_{\mathrm{PBH}} = \frac{\left<\sigma v\right>}{ m_\chi ^2}  4 \pi \int_0^{\infty}\rho(r)^2 r^2 dr \,,
\ee
where $\left<\sigma v\right>$ is the annihilation cross section and spherical symmetry of the density profile has been assumed. 
For a halo with a density profile
\be 
\rho(r)={\rm Min}(\rhom,\rhom(r/\rcut)^{-\alpha})\,,
\ee
and
assuming $\alpha>3/2$, the WIMP annihilation signal can be integrated into
\be 
\label{eq:Phi} 
\Gamma_{\mathrm{PBH}} = \frac{4 \pi \left<\sigma v\right> \rho^2_{\mathrm{max}} \rcut^3}{ m_\chi ^2}   \left( \frac{1}{3}+ \frac{1}{2\alpha-3}\right)\,,
\ee
where the first and the second terms in brackets are contributions from the constant-density central region and the falling profile, respectively. 
In the particular case of $\alpha=9/4$, this simplifies to 
\be 
\Gamma_{\mathrm{PBH}} = \frac{4 \pi \left<\sigma v\right> \rho^2_{\mathrm{max}} \rcut^3}{ m_\chi ^2}. 
\label{eq:GammaPBH}
\ee
In the above expression, a third of the contribution comes from the central region. Assuming a Heaviside density profile with the density dropping to zero at $\rcut$, as was done for example in \cite{Boucenna:2017ghj}, therefore underestimates the annihilation rate by a factor of $3$. For profiles that are less steep than $\alpha=9/4$ the contribution from the second term is even greater.

Equipped with these ingredients we can finally obtain the upper bound on the fraction of the dark matter in PBHs. For three different choices of the dark matter mass we find the constraint to be:
\begin{center}
\begin{tabular}{|c||c|c|c|}
\hline
$m_\chi$ & $10$ GeV  & $100$ GeV & $1$ TeV   \\
\hline
$f_{\mathrm{PBH}} \lesssim $ & $ 10^{-9}$ & $2 \times 10^{-9}$  & $4 \times 10^{-9}$  \\
\hline
\end{tabular}\label{fpbh-constraint}
\end{center}
This is a remarkably strong constraint which demonstrates that current limits on gamma-ray signals from WIMP annihilation imply that there are essentially no PBHs. On the other hand, a discovery of PBHs would imply extremely strong constraints on the amount of the dark matter in WIMPs, even if $f_{\mathrm{PBH}}$ were small. In that case the majority of dark matter would have to be some third species not considered here, such as axions.

This can be demonstrated as follows. 
Our derivation of the dark matter density profile around a PBH, Eq.~\eqref{eq:profile}, assumes that the only gravitational attraction on the dark matter particles is due to the PBH (i.e.~it neglects the effect of the dark matter which has already fallen into a halo around the PBH). This result therefore remains valid for the dark matter density profile of the WIMPs, provided the density is multiplied by $\fw\equiv\Omega_{\rm WIMP}/\Omega_{\rm DM}$, the fraction of dark matter in WIMPs. It then follows that the result for $r_{\rm cut}$ will be reduced by a factor of $\fw^{4/9}$ and hence $\Gamma_{\rm PBH}\propto\fw^{4/3}$, see Eqs.~\eqref{Eq:rcut} and \eqref{eq:GammaPBH}. 
Suppose
LIGO (or any other experiment) detected PBHs with density $f_{\rm PBH}=10^{-3}$, which is arguably the correct fraction to generate the black hole merger rate observed by LIGO and VIRGO, if one assumes most of the mergers are due to primordial rather than astrophysical black holes \cite{Sasaki:2016jop,Ali-Haimoud:2017rtz,Chen:2018czv,Raidal:2018bbj}. 
In this case the WIMP annihilation signal would  
be about a million times larger than the detected upper bound, assuming WIMPs formed the remainder of the dark matter. To make sure that the WIMP annihilation signal were acceptably small, the fraction of dark matter in WIMPs would have to be reduced to $\fw\lesssim 10^{-9/2} \simeq 3\times10^{-5}$.
Hence a detection of either WIMPs or PBHs would mean that the other component can form at most a tiny fraction of the dark matter.

Because $\Gamma_{\mathrm{PBH}}\propto M_{\rm PBH}$, the constraint on $f_{\mathrm{PBH}}$ is independent of the PBH mass, see Eq.~\eqref{fPBH}. This is true provided that the PBH mass is large enough to justify our approximation of neglecting the thermal kinetic energy of the WIMP particles compared to their gravitational potential energy at turn around. In particular, we need the kinetic energy to be small at the radius $r_{\rm cut}$ where the maximum density is reached. Even though their thermal energy will be larger at smaller radii, 
we expect this will just act to change the profile inside $r_{\rm cut}$ and not change the total mass of the WIMP particles inside the sphere of radius $r_{\rm cut}$. For PBHs with a mass comparable to those detectable by LIGO, a WIMP mass of $m_\chi=100$ GeV and $r_{\rm cut}\sim10^{-7} \kpc/h$ we can see from Fig.~\ref{fig:KEvsPE} that the kinetic energy is subdominant to the potential energy of the PBH by about five orders of magnitude,  and hence it should have a negligible effect. 
For much lighter PBHs the thermal kinetic energy is never negligible and our analytically derived profile cannot be used, invalidating our constraint. In that case, a more sophisticated treatment of the initial WIMP velocities along the lines described in \cite{Eroshenko:2016yve,Boucenna:2017ghj} should be made. 
Because the constraint on $f_{\mathrm{PBH}}$ is independent of the PBH mass, our constraint would also be valid if the PBH mass spectrum were not monochromatic, as long as most of the PBH masses were in the regime where the WIMP thermal kinetic energy is negligible.

We caution that the constraint on $f_{\rm PBH}$ was made assuming that the WIMP annihilation signal creates a diffuse gamma-ray background. Since the constraint on $f_{\rm PBH}$
is so tight, this will not be true for large PBH masses (because for fixed $f_{\rm PBH}$, the number density of PBHs is inversely proportional to their mass) and the constraint should be remade using the observational constraints from the Fermi satellite on point sources. However, doing so goes beyond the scope of this paper and there is no reason to expect the constraint to weaken by orders of magnitude.

\section{Conclusions}
\label{sec:conclusions}

We have demonstrated that WIMPs and PBHs are incompatible. If WIMPs make up the majority of the dark matter then $\fpbh\lesssim10^{-9}$, and if PBHs make up $1\%$ of the dark matter then WIMPs can only form about one millionth of the dark matter. These results are true for a broad range of WIMP and PBH masses, with the lower limit on the PBH mass being set by the thermal kinetic energy of the WIMPs. If this is large compared to the gravitational potential energy of the PBHs then high-density spikes around the PBHs will not form. 

The result that PBHs cannot coexist with WIMPs unless they form almost all or almost none of the dark matter was first reported in \cite{Lacki:2010zf}, who also found the constraints on $\fpbh$ to be independent of the PBH mass. However, they assumed an $r^{-3/2}$ density profile around the PBH, which we have shown is incorrect. The density profile close to the PBH (which is the relevant region for WIMP annihilation) is much steeper, being $r^{-9/4}$, as we have shown both analytically and numerically. 

Simulations of PBHs and dark matter (in a non-expanding background) were performed in \cite{Kavanagh:2018ggo}, to study the effect of the dark matter halos surrounding PBHs on the merger rate. However, they also assumed a $r^{-3/2}$ profile density around the PBH, and it would be interesting to repeat those simulations with an $r^{-9/4}$ density profile.

Our constraints are not valid for black holes which form through astrophysical processes since they form later when the background density of the Universe is much lower, and hence the DM halos around them would have a much lower maximum density. However, \cite{Bringmann:2009ip} shows that interesting constraints can be derived on intermediate-mass black holes even if they are not primordial.
We stress that the incompatibility of WIMPs and PBHs is only true for relatively massive PBHs. However, this range includes the entire mass range for which LIGO could detect a black hole merger, as well as heavier PBHs. At lower masses, the WIMPs' thermal kinetic energy becomes increasingly important compared to the gravitational potential energy of the PBH and our analysis breaks down. 
For a fiducial WIMP mass of $100$ GeV, this occurs at $\mpbh\sim 10^{-6}M_\odot$. The treatment of \cite{Eroshenko:2016yve}
and \cite{Boucenna:2017ghj} includes the WIMPs kinetic energy and they forecast 
that the existence of WIMPs would imply a relevant constraint (i.e.~$f_{\rm PBH}<1$) for PBHs with masses $\mpbh\sim 10^{-9}M_\odot$ (assuming a 70 GeV WIMP mass) and $\mpbh\sim 10^{-12}M_\odot$ (assuming a 100 GeV WIMP mass), respectively. Therefore a detection of WIMPs would not rule out the existence of light PBHs or PBH relics, but it would rule out the possibility that LIGO has detected PBHs as well as the possibility that supermassive black holes have primordial seeds \cite{Bernal:2017nec}.

\acknowledgments

We thank   Sofiane Boucenna, Neal Dalal, Yury Eroshenko, Oliver Hahn, Bradley Kavanagh, Florian K\"{u}hnel, Go Ogiya, and Luca Visinelli for useful discussions.  JA is supported by STFC Consolidated Grant
ST/P000592/1. CB is supported by a Royal Society University Research Fellowship. MG acknowledges support from the Marsden Fund of the Royal Society of New Zealand and from the European Research Council under the European Union's Seventh Framework Programme  (FP/2007-2013)/ERC Grant Agreement No.~616170.
This work was supported by a grant from the Swiss National Supercomputing Centre (CSCS) under project ID s710.

\appendix

\section{A more accurate derivation of the halo profile}\label{app:halo}

In the main text we derived the dark matter density profile around a PBH assuming that dark matter particles ``freeze-in'' when they decouple from the background expansion. Even neglecting their kinetic energy this is a poor approximation, since they would instead oscillate on a radial orbit centred on the PBH (in reality a highly elliptical orbit, with most WIMPs narrowly passing but not falling into the PBH). In that case, the halo density is given by
\be \rho(r)=\frac{1}{r^2}\int_r^\infty {\rm d}r_i\, r_i^2 \rho(r_i,t_i)\frac{2}{T_{\rm orbit}}\frac{{\rm d}t(r_i)}{{\rm d}r},
\ee
where $r_i$ is the initial radius of the WIMP orbits at the time they decouple from the background expansion (and $t_i$ the corresponding time), $T_{\rm orbit}=\pi r_i^{3/2}r_g^{1/2}$ is the period of the orbit and $\rm{d}t/\rm{d}r=r_g^{-1/2}(1/r-1/r_i)^{-1/2}$ describes the time dependence of the orbit and the gravitational radius is $r_g^2=2G\mpbh$ \cite{Eroshenko:2016yve,Boucenna:2017ghj}. 
The integral is over $r>r_i$ because particles with negligible initial kinetic energy never move a larger distance from the PBH than their initial separation when they first decouple from the background expansion. Performing the integral produces a profile with an $r^{-9/4}$ profile like \eqref{eq:profile} but which is $53\%$ denser. Because the integral above should be truncated at matter-radiation equality rather than $\infty$, the size of this correction of including the radial motion of the particles in the halo will in reality be smaller.

\section{Numerical solutions of the turnaround radius.}
\label{app:numprof}

In this appendix we present a numerical solution to equation \eqref{eq:r-eom} and use that solution to justify equation \eqref{eq:rta}. In radiation domination $\ddot{a}/a = -1/(4t^2)$, so equation \eqref{eq:r-eom} can be rewritten as
\be
\ddot{r} = -\frac{GM_{\rm PBH}}{r^2} - \frac{r}{4t^2}\,.
\ee

This can be written in a dimensionless form via the transformations $y=r/(2GM_{\rm PBH})$ and $\tau=t/(2GM_{\rm PBH})$, giving
\be
\ddot{y} = -\frac{1}{2y^2}- \frac{y}{4\tau^2}.
\label{eq:y-eom}
\ee
In equation \eqref{eq:y-eom} the overdot refers to differentiation with respect to $\tau$ rather than $t$. Because the differential equation can be rewritten in this form, independent of the PBH mass $M_{\rm PBH}$, a family of solutions found for one mass can be rescaled onto the solutions for other masses.

At early times the cosmological background solution should dominate. Therefore, the initial time $\tau_0$ and radius $y_0$ for the numerical solution are set so that $y_0/(4\tau_0^2)\gg 1/(2y_0^2)$. $\dot{y}$ is then determined by the cosmological solution during radiation domination, i.e. $r \propto t^{1/2}$. This gives $\dot{y_0}=y_0/(2\tau_0)$.

\begin{figure}
	\begin{center}
		\includegraphics[scale=0.6]{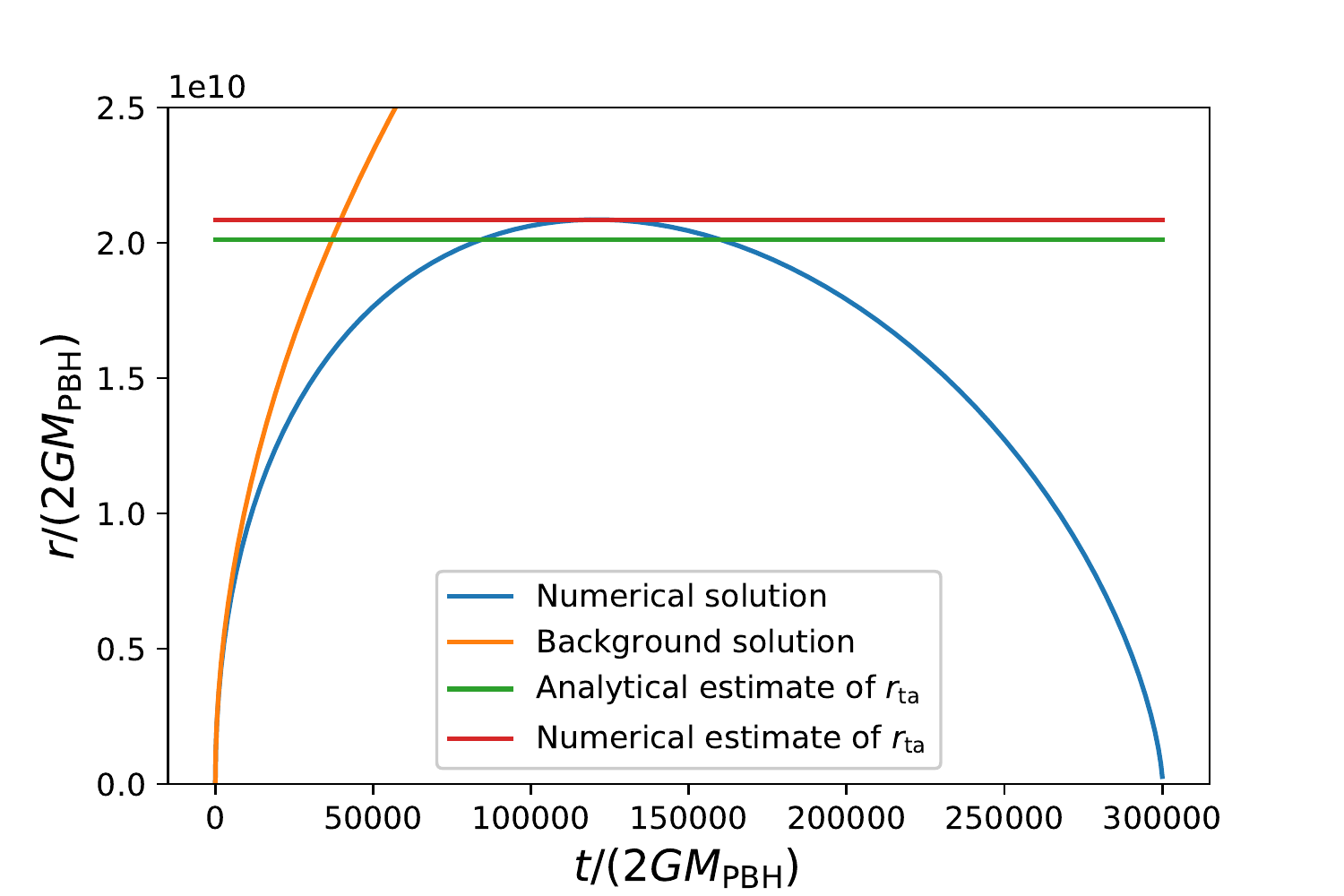}
	\end{center}
	\caption{The rescaled radius as a function of rescaled time for a spherical shell of particles around a PBH during radiation domination. We plot the full numerical solution, the background solution and two estimates of the turnaround radius. The numerical estimate of the turnaround radius in equation \eqref{eq:rta} clearly works very well.}
	\label{fig:numtraj}
\end{figure}

In figure \ref{fig:numtraj} we show a comparison of a numerical solution of equation \eqref{eq:y-eom} to the background cosmological solution. We also plot two horizontal lines showing our analytical and numerical estimates of the turnaround radius (i.e.~equations \eqref{eq:ta-analytic} and \eqref{eq:rta}). It is clear that the analytical estimate does surprisingly well, but that the numerical estimate does better. In fact, the ratios of the two estimated turnaround radii to the actual turnaround radius are 0.964 and 0.999 for what we have called the analytical and numerical estimates, respectively. This result is what motivates us to use equation \eqref{eq:rta} for our main results. If there is an analytical way of deriving this result we have been unable to find it. 

\bibliographystyle{JHEP}
\bibliography{1-bibfile}

\end{document}